# AI Spillover is Different: Flat and Lean Firms as Engines of AI Diffusion and Productivity Gain

Gavin Wang        Chun Feng        Tianshu Sun
UT Dallas         CKGSB            CKGSB

*This version: 2025.11.02*

## Abstract

Labor mobility is a critical source of technology acquisition for firms. This paper examines how artificial intelligence (AI) knowledge is disseminated across firms through labor mobility and identifies the organizational conditions that facilitate productive spillovers. Using a comprehensive dataset of over 460 million job records from Revelio Labs (2010–2023), we construct an inter-firm mobility network of AI workers among over 16,000 U.S. companies. Estimating a Cobb–Douglas production function, we find that firms benefit substantially from the AI investments of other firms from which they hire AI talents, with productivity spillovers two to three times larger than those associated with traditional IT after accounting for labor scale. Importantly, these spillovers are contingent on organizational context: hiring from flatter and more lean-startup-method-intensive firms generates significant productivity gains, whereas hiring from firms lacking these traits yields little benefit. Mechanism tests indicate that "flat and lean" organizations cultivate more versatile AI generalists who transfer richer knowledge across firms. These findings reveal that AI spillovers differ fundamentally from traditional IT spillovers: while IT spillovers primarily arise from scale and process standardization, AI spillovers critically depend on the experimental and integrative environments in which AI knowledge is produced. Together, these results underscore the importance of considering both labor mobility and organizational context in understanding the full impact of AI-driven productivity spillovers.

# 1. Introduction and Motivation

The rise of artificial intelligence (AI) is reshaping the global economic landscape at an unprecedented pace, sparking a surge of worldwide investment as firms commit resources at record levels to secure a competitive edge in the AI era (Chen and Chan, 2024; Babina et al., 2024; McElheran et al., 2024). Global corporate AI investment reached $252.3 billion in 2024, a 25.5% increase over 2023, and the share of surveyed firms reporting the use of some form of AI has jumped from 55% in 2023 to 78% in 2024[1]. This wave of capital investment has ignited an intense talent war. Acquiring top-tier AI talent has emerged as a core strategic battleground for firms competing for future market leadership. For example, in 2025, Meta reportedly offered compensation packages ranging from $200 million to $500 million to recruit leading AI researchers, illustrating the extraordinary intensity of the AI talent market.[2]

Despite this war for AI talent and the rapid movement of skilled workers across firms, little is known about whether and how AI-related knowledge spills over through labor mobility. This gap is particularly important because hiring AI professionals does not automatically embed their expertise into organizational processes or translate into productivity gains—the current "AI productivity paradox" (Brynjolfsson et al., 2021; McElheran et al., 2025). The challenge is evident even among leading technology firms: in October 2025, Meta laid off more than 600 employees from its AI division, highlighting that the presence of top AI talent alone does not ensure that their knowledge becomes organizational capability.³ Understanding the mechanisms and directions of AI knowledge spillovers is therefore not only crucial for capturing value from AI talent but also for informing firms' strategic decisions on *where* to recruit such talent. Moreover, the nature of AI capabilities differs fundamentally from earlier waves of IT. Prior research indicates that IT-related knowledge often disseminates through worker mobility, typically from large firms that possess codified routines and standardized workflows, enabling digital technologies to scale (Bresnahan et al., 2002; Aral et al., 2012; Tambe and Hitt, 2014; Wu et al., 2018). By contrast, modern AI systems—characterized by high predictive power but low interpretability—require continuous experimentation and iterative fine-tuning (Berente et al., 2021; Wang and Wu, 2024; Tambe, 2025). These characteristics make the flow and diffusion of AI knowledge more reliant on adaptive, fast-paced environments, shifting the potential locus of spillovers away from large-firm settings toward more experimental organizational contexts.

---

[1] Data Source: Stanford University AI Index Report 2025: https://hai.stanford.edu/ai-index/2025-ai-index-report.
[2] Source: https://www.businessinsider.com/meta-escalates-ai-talent-war-with-openai-shengjia-zhao-zuckerberg-2025-7.

In this study, we investigate the productivity spillovers generated by inter-firm mobility of AI talents and how the organizational characteristics of source firms moderate these effects. Using detailed labor mobility data from Revelio Labs spanning over 460 million job records from 2010 to 2023, we construct a directed network of AI talent flows across 16,000 U.S. firms. Extending prior work on IT spillovers (e.g., Tambe and Hitt, 2014; Wu et al., 2018), we distinguish not only whether firms gain from hiring AI workers, but also how the source firms' internal practices, namely, hierarchical flatness and the use of the Lean Startup Method (LSM)—a development approach emphasizing rapid experimentation, iterative development, and data-driven decision-making, mediate the spillovers. We focus on flat and lean organizational features because modern AI work itself is highly experimental, iterative, and data-driven, requiring flexible structures to translate individual expertise into organizational learning (Berente et al., 2021; Wang and Wu, 2024; Tambe, 2025). We find that recipient firms reap the most significant benefits when hiring AI workers from flatter, experimentation-oriented organizations. By contrast, source firms without these organizational features generate weaker or no spillover benefits for their talent recipients. This evidence highlights that the nature of AI spillovers is not purely technological but fundamentally organizational—rooted in the environments where AI talent previously operated.

---

³ Source: https://www.reuters.com/business/meta-is-cutting-around-600-roles-ai-unit-axios-reports-2025-10-22/.

Mechanism tests indicate that "flat and lean" organizations cultivate more versatile AI generalists who transfer richer knowledge across firms. Acquiring talents from such source firms yields higher productivity and better innovation performance. Comparing AI spillovers with AI-excluded IT spillovers, we find that AI productivity spillovers are two to three times larger than those associated with traditional IT after accounting for labor scale. These results remain robust to a variety of instrumental variable strategies, including leveraging exogenous shocks to one's hiring network structures and state-level enforcement of non-compete agreements.

This study contributes to the information systems (IS) literature by deepening our understanding of digital capability spillovers in the AI era. While prior IS research has established the productivity effects of IT labor and the diffusion of digital practices through labor mobility, relatively little attention has been paid to how organizational practices at source firms moderate the spillover effects on recipient firms. By operationalizing novel measures of LSM adoption and hierarchical flatness and embedding them within a labor mobility network, we offer a fresh lens on knowledge spillovers and organizational complementarities. Our findings also contribute to explaining the current "AI productivity paradox" (Brynjolfsson et al., 2021; McElheran et al., 2025) and inform broader IS debates on whether AI knowledge diffusion demands distinct organizational strategies compared to prior digital transformation waves, suggesting that AI represents not only a new class of technology but also a fundamentally different model of capability diffusion.

## 2. Literature and Theory

In this section, we first review the growing body of research examining the impact of AI on firm productivity and how it relates to traditional IT. We then build on these insights to theorize how AI-related knowledge diffuses across firms and analyze how talent mobility contributes to productivity spillovers.

**2.1 AI and Firm Productivity**

In this study, we align with recent literature defining AI as advanced computing power for making predictive actions using machine learning techniques, often emphasizing high accuracy at the expense of interpretability (Heath, 2019; Taddy, 2019; Berente et al., 2021; Brynjolfsson et al., 2021). The "black-box" nature of AI algorithms represents a key difference from traditional IT and highlights the importance of understanding the context of the data being used (Taddy, 2019; Berente et al., 2021; Li et al., 2021). A growing body of scholarship has examined the impact of AI on individuals and firms' innovation capabilities and operational efficiency (Verganti et al., 2020; Csaszar and Steinberger, 2022; Babina et al., 2024; Schanke et al., 2024). For example, Acemoglu and Restrepo (2020) highlight AI's ability to automate routine human tasks, thereby yielding substantial savings in labor expenditures. Similarly, Shrestha (2021) emphasizes that AI, through more accurate data analytics and predictive modeling, enables firms to optimize operational decision-making and elevate overall performance. Moreover, recent evidence from Babina et al. (2024), based on publicly listed U.S. firms, demonstrates that active AI adopters not only exhibit superior sales growth and market valuations but also achieve greater capacity for new product development, driving the expansion of their operational scale.

While AI carries substantial potential to foster innovation and productivity, a number of scholars observe that many firms that have invested heavily in the technology have failed to realize commensurate returns, the so-called " AI productivity paradox" (Brynjolfsson et al., 2021; McElheran et al., 2025; Wang and Wu, 2024). To address this puzzle, Brynjolfsson et al. (2021) propose the now well-cited "productivity J-curve," which posits that firms adopting AI often experience a temporary dip in efficiency during the initial phases, with productivity gains materializing only after a period of adjustment, capability building, and organizational restructuring. Supporting this perspective, McElheran et al. (2025), drawing on an analysis of U.S. manufacturing firms, show that enterprises deploying industrial AI frequently encounter short-term bottlenecks stemming from misalignments between new AI tools and pre-existing operational practices. However, as these firms systematically invest in complementary assets—including data infrastructure, employee reskilling, and process reengineering—their productivity trajectories eventually shift from negative to positive. Collectively, these findings underscore that the impact of AI on firm productivity is highly contingent upon the degree of organizational transformation and process innovation accompanying its adoption. Simply introducing AI is insufficient to ensure performance improvements.

However, even for firms that invest in organizational transformation, translating AI capabilities into sustained performance gains remains a complex endeavor. This complexity arises in part because AI expertise is deeply embedded in human capital—particularly in the tacit knowledge, modeling heuristics, and domain experience of skilled professionals (Tambe, 2025). As a result, the movement of such talent across firms may represent an important, yet underexplored, mechanism through which AI-related knowledge and capabilities diffuse across the economy.

**2.2 Talent Flow and Technological Spillovers**

Technological "spillovers" refer to instances where the benefits of a firm's innovation and technology investments extend beyond its boundaries, allowing other firms—often without equivalent investments—to capture value indirectly (Griliches, 1991; Bresnahan et al., 2002). Extensive empirical research supports the existence of such spillovers, with employee mobility across organizations widely regarded as a critical conduit for the diffusion of technical knowledge (Arrow, 1962; Fallick et al., 2006). This is especially important for IT, because the technical know-how required to implement new IT innovations is primarily embodied within the IT workforce (Tambe and Hitt, 2014; Dedrick et al., 2003; Draca et al., 2006; Oettl and Agrawal, 2008). For example, Tambe and Hitt (2014) estimate that firms hiring experienced IT workers from other IT-intensive organizations can capture productivity gains equivalent to 20%–30% of the originating firms' own IT investment returns. Building on this perspective, Wu et al. (2018) argue that firms sourcing talent from structurally diverse networks can access a broader array of non-redundant, innovative insights. These findings suggest that the diffusion of technical knowledge through labor mobility is not a random or passive process. Instead, it depends critically on firms' ability to identify high-value talent pools—particularly those embedded in organizations with advanced technological capabilities—and to successfully transfer, absorb, and integrate the embodied knowledge these workers bring.

While much of the prior literature has focused on traditional IT, recent developments suggest that AI may generate spillovers through a structurally similar mechanism. Like IT, AI adoption requires specialized technical expertise—such as algorithm development, data engineering, and system integration—that is often tacit and embedded in individual workers (Wang and Wu, 2024; Tambe, 2025). These skills are typically acquired through experiential learning in high-capability environments, making the movement of AI professionals across firms a potentially powerful channel for knowledge diffusion.

Moreover, as with IT, many firms seeking to adopt AI lack the in-house capability to fully exploit the technology, leading them to hire from external labor markets (Babina et al., 2024). In this context, labor mobility enables firms to indirectly access the technological insights, tooling practices, and implementation heuristics developed by frontier firms. Although the nature of AI expertise may be more context-specific than traditional IT, the basic logic of skill transfer through human capital remains applicable. If anything, the rapid pace of AI advancement and the scarcity of experienced talent may amplify the strategic value of hiring from high-investment firms.

Therefore, we hypothesize that AI investments generate productivity spillovers through AI labor flows among firms:

***H1:*** *The AI investments of other firms generate productivity spillovers through AI labor flows among firms.*

**2.3 AI is Different: Standardized versus Adaptive Environments**

Prior studies have shown that traditional IT-related spillovers typically originate from large, established organizations with intensive IT investments, as the codified processes and standardized workflows in such firms make technical knowledge more transferable and easier to replicate in other settings (Bresnahan et al., 2002; Aral et al., 2012; Tambe and Hitt, 2014; Cao et al., 2025). In these contexts, IT systems—such as enterprise resource planning (ERP) or customer relationship management (CRM)—are often deployed following well-documented implementation protocols, industry-wide best practices, and vendor-provided training modules. These features render the underlying knowledge relatively explicit, modular, and detached from firm-specific routines, thereby facilitating its migration through technical manuals, consultants, and employee movement. As a result, downstream firms can adopt and benefit from IT innovations with comparatively low absorptive barriers.

However, the same logic may not apply to AI, where knowledge is often context-dependent and embedded in complex human–machine interactions due to the "black-box" nature of AI algorithms, potentially shifting the locus of spillovers toward more adaptive and experimental organizational environments (He et al., 2020; Wang and Wu, 2024). In contrast to traditional IT systems—whose adoption can follow standardized implementation protocols or industry-wide best practices—AI technologies typically require close integration with domain-specific processes, datasets, and contexts. The effectiveness of AI thus hinges not only on technical sophistication, but also on the ability to align models with the nuanced operational realities in which they are deployed (Berente et al., 2021; van den Broek, 2025). This alignment demands contextual knowledge, such as an understanding of organizational workflows, data-generation processes, user incentives, and feedback dynamics—all of which are rarely codified and difficult to abstract across firms. For example, engineers at Netflix have noted that AI success depends not just on modeling capabilities, but also on a deep familiarity with platform-specific user behaviors and KPIs, cultivated through continuous experimentation and organizational feedback loops.[4]

Crucially, this contextual knowledge is not easily learned in top-down, rigidly structured organizations where AI development is siloed or disconnected from frontline operations. Instead, it is more likely to emerge in adaptive environments that promote iterative experimentation, decentralized decision-making, and close collaboration between AI developers and domain experts (Tursunbayeva and Chalutz-Ben Gal, 2024; Wang and Wu, 2024). In such settings, AI professionals are continually exposed to real-world deployment challenges, enabling them to internalize the organizational and behavioral conditions that shape AI performance. Without this embedded experience, even technically skilled AI workers may struggle to adapt their knowledge to new firm settings—limiting the extent to which labor mobility generates productivity spillovers.

This suggests that the source environment in which AI talent is cultivated plays a critical role in shaping whether—and how—their knowledge can be transferred across firms.

*2.3.1 Lean Startup Method (LSM) and Contextual Knowledge*

---

[4] Source: https://netflixtechblog.com/foundation-model-for-personalized-recommendation-1a0bd8e02d39.

Originating from the work of Blank (2003) and Ries (2011), LSM advocates that firms should validate business ideas through minimum viable products (MVPs), iterative build–measure–learn cycles, and market-driven experimentation to achieve product innovation (Rigby et al., 2018). Unlike the top-down, strategy-driven innovation mode, LSM advocates a bottom-up, experimentation-driven agile approach. Firms typically achieve LSM by rapidly and iteratively using prototyping (e.g., building MVPs or prototypes) and controlled experimentation (e.g., A/B testing) for innovation (Yoo et al., 2021; Koning et al., 2022; Wang and Wu, 2024).

With its emphasis on rapid experimentation, user-centric iteration, and cross-functional learning, LSM offers an ideal organizational environment for cultivating contextual knowledge that complements AI capabilities. Unlike top-down innovation processes that separate technical teams from operational feedback, LSM embeds AI engineers directly in cycles of build–measure–learn, forcing them to confront real-world constraints, data frictions, and user behaviors. This close coupling of technical design with practical implementation enables AI talent to internalize what works, what breaks, and how to recalibrate—insights that are not captured in models or documentation, but are essential for successful knowledge transfer across firms.

For example, at Airbnb, engineers working on AI-driven search and pricing algorithms operate within fast-paced experimentation environments, where A/B tests are launched weekly and cross-functional teams iterate rapidly based on user behavior and platform dynamics.[5] This fosters not only technical proficiency but also a practical understanding of how to align AI outputs with business KPIs and customer expectations—insights that accompany employees when they transition to other firms. In contrast, AI professionals coming from more centralized, strategy-driven firms may lack this embedded operational fluency and struggle to adapt their skills to new organizational contexts.

---

[5] Source: https://airbnb.tech/data/beyond-a-b-test-speeding-up-airbnb-search-ranking-experimentation-through-interleaving/.

Similarly, firms like ByteDance have cultivated a culture of high-frequency experimentation and decentralized decision-making, where AI teams are empowered to iterate on recommendation algorithms in direct response to granular user data.[6] Engineers trained in such LSM-intensive environments accumulate not only algorithmic expertise but also adaptive heuristics for applying AI in complex, fast-changing domains like content moderation and personalized advertising—skills that prove highly valuable when transferred to new firms.

These examples underscore the idea that the source environment in which AI talent is trained fundamentally shapes the portability and productivity of their knowledge. Thus, when AI talent is sourced from organizations with stronger LSM capabilities, they are more likely to carry transferable deployment knowledge and practical heuristics that can be readily adapted to new contexts. In contrast, AI workers from less experimentation-driven firms may possess strong technical foundations but lack the embedded operational experience required to translate those skills into productivity gains in a different organizational setting.

We therefore hypothesize that:

*H2: All else equal, firms hiring AI talents from source firms with higher LSM capabilities are more*

---

[6] Source: https://jefftowson.com/membership_content/bytedance-is-going-for-fast-decentralized-innovation-at-global-scale-tech-strategy-daily-article/ .

*productive than firms hiring AI talents from source firms with lower LSM capabilities.*

### 2.3.2 Flat Organizational Structures and AI Knowledge Spillovers

Beyond experimentation capabilities, the organizational hierarchy of the source firm can also shape the portability of AI-related knowledge. **Flat organizations**, characterized by fewer managerial layers and greater employee autonomy, enhance horizontal communication and cross-functional collaboration (Blau and Scott, 2003; Burton and Obel, 2004). By minimizing rigid reporting chains and standardized routines, such structures facilitate real-time information exchange and lateral idea generation, enabling employees to self-organize around emerging problems and solutions (Puranam and Håkonsson, 2015; Lee and Edmondson, 2017). These environments are particularly conducive to innovation, as they support the cross-pollination of diverse perspectives and reduce internal knowledge silos (Saxenian, 1996; Fleming et al., 2007). While flat structures and lean startup practices often co-occur and reinforce one another, they are conceptually distinct: LSM emphasizes iterative experimentation and customer validation, whereas flat structures refer to the formal reduction of hierarchical layers that shape communication and decision-making authority.

The structural advantage of flatness is especially relevant to AI development, which relies heavily on cross-disciplinary coordination, fast iteration, and context-aware deployment. As Puranam (2021) notes, the rise of human–machine collaboration demands flatter organizational designs that enable agile workflows and decentralized decision-making. Reitzig (2022) and Snow et al. (2017) similarly argue that flattening hierarchies accelerates interdepartmental coordination and facilitates the integration of AI technologies into business functions. Case-based evidence also supports this view: Adobe's Kickbox program demonstrates how empowering employees through a hybrid model of flexible hierarchy and distributed authority can enhance the translation of AI-driven insights into deployable innovations (Baumann and Wu, 2023).

In contrast, AI professionals from rigidly hierarchical firms often operate in narrowly scoped roles with limited exposure to deployment contexts or user feedback. While technically capable, they may lack the embedded operational insight needed to adapt their knowledge across organizations. By comparison, individuals trained in flatter structures tend to be more adept at transferring tacit knowledge, collaborating across functions, and rapidly integrating their expertise into new environments. These qualities reduce adaptation costs and accelerate impact, increasing the likelihood of productivity-enhancing spillovers following interfirm mobility.

We therefore hypothesize:

***H3:*** *All else equal, firms hiring AI talents from source firms with flatter hierarchies are more productive than firms hiring AI talents from source firms with higher hierarchies.*

## 3. Data and Setting

### 3.1 General Data Description

Our study relies on comprehensive U.S. workforce mobility data from 2010 to 2023, sourced from the Revelio Labs database, which provides structured records of individuals' employment histories, including employer names, job titles or descriptions, and start and end dates for each position. It has been widely used in labor mobility research (Baker et al., 2024; Tambe, 2025). Our final dataset comprises approximately 460 million job records in total. To identify AI talents, we construct a keyword dictionary based on the AI-related skills lists (Table 1) provided by Babina et al. (2024) and Wang and Wu (2024). Any talent whose job title or description contains one or more of these keywords is classified as an AI talent.[7] In total, we identify approximately 1.2 million AI talents, accounting for about 0.2% of all talents in the dataset, a proportion consistent with that reported by Babina et al. (2024). Based on this identification, we further construct an inter-firm AI talent mobility network. Specifically, we extract the employer names and start and end dates for each AI talent's job history and trace their employment transitions chronologically. Treating firms as nodes and inter-firm job changes as directed edges, we follow Wu et al. (2018) and set a 5-year mobility window: if a focal (recipient) firm hires at least one AI talent from another firm within the past five years, we draw a directed edge between the two firms. Edge weights follow Tambe and Hitt (2014) and are defined as the number of talents transitioning along each path.

<Table 1 Inserted Here>

We then match these firms to the Compustat database of public companies and aggregate the data at the firm-year level. This process yields a sample of 3,502 U.S. public firms, among which 2,189 firms have hired AI talents between 2010 and 2023, sourced from 15,964 origin firms, resulting in a network of 109,045 directed inter-firm job transitions. Figure 1 visualizes the AI talent flow networks in 2023 for the top recipient firms, including Amazon, Microsoft, Apple, IBM,

---

[7] To improve the accuracy of AI skill identification, we apply rigorous semantic disambiguation and filtering to employees with abbreviations such as "AI," "ML," "DL," and "CV." Specifically, we exclude Latin-based terms where "ai" functions only as a common auxiliary or pronoun (e.g., the French "j'ai"), unrelated to artificial intelligence, and manually remove non-AI meanings of abbreviations (e.g., "CV" frequently referring to "curriculum vitae"). These steps eliminate 135,550 employee records (approximately 9.5% of the original sample), ensuring the accuracy and robustness of subsequent AI talent measures.

and Alphabet.

<Figure 1 Inserted Here>

**3.2 Variables**

*Measuring AI Spillover*

We follow the approach in Tambe and Hitt (2014) and Wu et al. (2018) to construct the AI spillover variable: *AI pool*. We adapt their method by replacing IT talents with AI talents. Specifically, the *AI pool* variable represents the weighted average AI talent intensity among the source firms that have contributed AI talents to the focal firm (recipient), where AI talent intensity is defined as the proportion of AI talents in a given source firm's total workforce, and is weighted by the share of AI talents that each source firm contributes to the focal firm:

$$AI\,pool = \sum_{i \in \{Hiring\,Sources\}} AI\,Talent\,Intensity_i \times p_i$$

$$AI\,Talent\,Intensity_i = \frac{AI\,labors_i}{total\,labors_i}\,;\ p_i = \frac{inflow_i}{\sum_{k \in \{Hiring\,sources\}} inflow_k}$$

*Measuring LSM*

We measure the extent to which a startup adopts LSM based on job descriptions of current employees, obtained from Revelio Labs. These job roles reflect the company's existing strategic orientation and operational focus, providing insight into how actively the company applies LSM to drive product innovation. We follow the current literature and count the number of employees whose job titles and job descriptions contain keywords commonly related to LSM (Table 1) as the firm's LSM level each year (Wang and Wu, 2024). These keywords are extracted from descriptions of job roles, capturing main activities in LSM-related practice, including Prototyping (e.g., trial product, prototype development) and Controlled Experimentation (e.g., A/B testing, experimentation). Consistently, we use keyword-based methods to separately capture the extent to which each firm engages in these two types of activities. Companies that have or hire employees with expertise in such activities are more likely to have adopted LSM than others. Similarly, we also create the metric *LSM AI pool* to capture the weighted average LSM talent intensity among the source firms that have contributed AI talents to the focal firm (recipient). The LSM talent intensity of a source firm is calculated as the proportion of LSM talents in a given source firm's total workforce.

$$LSM\ AI\ pool = \sum_{i \in \{Hiring\ sources\}} LSM\ Talent\ Intensity_i * p_i$$

$$LSM\ Talent\ Intensity_i = \frac{LSM\ labors_i}{total\ labors_i};\ p_i = \frac{inflow_i}{\sum_{k \in \{Hiring\ sources\}} inflow_k}$$

*Measuring Hierarchy/Flatness*

We first obtain the number of distinct hierarchical levels within a company, following the approach developed by Lee (2022). Specifically, we classify each employee's job title into one of 12 predefined levels (i.e. "Owner," "President," "VP," "CEO," "C-Suite," "Head," "Director," "Manager," "Producer," "Lead," "Supervisor," and "Other") and then count the number of unique levels present in each firm (Table 1). This method draws on Lee's validated framework, which relies on practitioner-verified keyword classifications to infer organizational structure from job titles. To construct the hierarchy variable for empirical analysis, we classify each employee into one of the 12 levels based on whether their job title contains specific keywords. When a job title contains keywords corresponding to two or more levels, the employee is categorized under the highest-ranking level. The final hierarchy measure is the number of unique levels with at least one employee in the firm.

To test our hypothesis that firms enjoy more spillovers when they hire AI talents from companies with flatter hierarchical structures, we adapt the *AI pool* metric following the practice in Wu et al. (2018) and create a new metric, *flat AI pool*, capturing the weighted average hierarchical flatness among the source firms that have contributed AI employees to the focal firm (recipient). The hierarchical flatness of a source firm is calculated as the ratio of the total number of employees to the total number of hierarchical levels, as a company is considered flatter if it has fewer hierarchical levels, given the same number of employees.

$$Flat\ AI\ pool = \sum_{i \in \{Hiring\ sources\}} Hierarchical\ Flatness_i * p_i$$

$$Hierarchical\ Flatness_i = \frac{total\ employees_i}{number\ of\ hierarchical\ levels_i}; \quad p_i = \frac{inflow_i}{\sum_{k \in \{Hiring\ sources\}} inflow_k}$$

There might be concerns that firms exhibit heterogeneous hierarchical structures across departments, potentially introducing measurement noise to our firm-level flatness metric. To examine this issue, we calculate the within-firm variation in hierarchy levels across departments for all source firms. As shown in Appendix A, the largest departmental differences in hierarchy levels average about one level and are statistically insignificant (p-value> 0.8). Hence, the firm-level flatness measure used in our analysis is unlikely to be distorted by within-firm heterogeneity in organizational hierarchy.

Following Tambe and Hitt (2014) and Wu et al. (2018), we impute missing observations with industry averages. Our final panel data comprises 49,027 observations for 3,502 U.S. public firms spanning the period from 2010 to 2023. Table 2 shows summary statistics.

<Table 2 Inserted Here>

## 4. Empirical Methodology

## 4.1 Base Model

Following the classic methods used in information systems and productivity literature (Bresnahan, 2003; Tambe and Hitt, 2014; Wu et al., 2018), we employ a Cobb-Douglas production function framework, relating firms' output (*Sales*) to primary firm-level control variables, materials (*M*), Capital (*K*), AI labor (*L_AI*), non-AI labor (*L_non-AI*), and measures of spillover effects (*AI pool, Flat AI pool, LSM AI pool*). All variables are log-transformed to capture the factor share of their productivity contribution,[8] and the core regression specification is as follows:

$$\log Sales_{i,t} = \beta_1 \log M_{i,t} + \beta_2 \log K_{i,t} + \beta_3 \log L\_non\_AI_{i,t} + \beta_4 \log L\_AI_{i,t} + \beta_5 \log AI\_Pool_{i,t}$$
$$+ \beta_6 \log Flat\_AI\_Pool_{i,t} + \beta_7 \log LSM\_AI\_Pool_{i,t} + \delta_i + \lambda_t + Controls + \varepsilon_{i,t}$$

We include year fixed effects ($\lambda_t$) in all regressions to control for time-specific productivity shocks and industry fixed effects ($\delta_i$) using 2-digit NAICS code to address industry heterogeneity. Following the IT spillover literature, we control for industry rather than firm fixed effects. This approach captures within-industry productivity spillovers identified in prior studies while avoiding collinearity problems that arise because firms' hiring sources exhibit limited temporal variation (Chang and Gurbaxani, 2012; Tambe and Hitt, 2014; Wu et al., 2018).

---

[8] When applying the logarithmic transformation, this study does not adopt the conventional "add-one" adjustment (i.e., log(x + 1)) commonly used to address the issue of zeros in logarithms. This is because the *AI pool, Flat AI pool, LSM AI pool* variables contain a large number of zeros in the raw data and the non-zero observations are generally small (e.g., 0.001). Directly adding one before taking logarithms would substantially amplify the disparities in the data distribution and introduce systematic bias. To more appropriately handle such data characteristics, we follow the approach of Stahel (2002) and apply a log(x + c) transformation, where c is set to one-half of the smallest strictly positive value of the variable (i.e. c = min(x>0)/2). This method retains the information carried by zero observations while minimizing distortion of the original distribution, making it suitable for logarithmic transformations of sparse, small-valued continuous variables.

## 4.2 Identification Strategy

Although the AI intensity, organizational hierarchies, and LSM intensity of source companies are exogenous to the focal firm, endogeneity concerns may still arise in this study because the focal firm endogenously makes the hiring decisions. Higher-quality companies may choose to hire talent from more AI-intensive companies, thereby biasing the AI pool estimate upward. Although offering conclusive causal evidence proves challenging due to the endogeneity of organizational practices in observational data, we try to alleviate such concerns through two sets of instrumental variables that generate exogenous shocks from different dimensions.

The first set of instruments is based on the enforcement level of non-competing agreements (NCA) in different states. NCA refers to restrictive clauses signed between employees and employers, under which employees are prohibited from joining competitor firms or starting their own businesses for a certain period after departure (Shi, 2023). Prior studies have widely documented that NCAs, through legal enforcement, significantly suppress employees' horizontal mobility and job-switching behavior (Garmaise, 2011; Marx et al., 2009; Starr et al., 2018). Shi (2023) notes that firms utilize NCAs to capture employees' future matching rents, thereby raising the barriers to job transitions and constraining talent mobility. Empirical evidence shows that executives who sign NCAs experience an approximately 1.8% lower annual turnover probability, confirming the tangible restrictive effect of such institutional arrangements on talent mobility.

Although many states prohibit the signing of non-compete agreements, the enforcement intensity of NCAs varies substantially across states and over time within states. This enforcement intensity is exogenous to firms' internal hiring decision-making and exerts a significant influence on their hiring behavior (Starr et al., 2018; Shi, 2023). For example, California's outright ban on NCAs has substantially increased employee mobility in the information technology sector, facilitating the formation of high-tech clusters such as Silicon Valley (Fallick et al., 2006).

We follow recent literature in calculating the enforcement level of NCA in each state each year (Shi, 2023), and construct a weighted average NCA enforcement intensity for each firm each year based on the source states of AI talent inflow of this firm:

$$NCA = \sum_{i \in \{Hiring\ sources\}} state\_NCA\_level_i * p_i$$

$$p_i = \frac{inflow_i}{\sum_{k \in \{Hiring\ sources\}} inflow_k}$$

This variable captures the external labor mobility regulatory environment associated with the focal firm's hiring activities. Its theoretical basis lies in the fact that firms often exhibit geographic preferences in their hiring (Lin and Viswanathan, 2016; Chan and Wang, 2018), including AI talent (Babina et al., 2024). As a result, the shifts in NCA enforcement levels in these states may affect the mobility and employability of potential candidates, thereby influencing the flow of AI talents. However, such NCA-related institutional arrangements and constraints are exogenous to a company's unobserved qualities, especially considering that a substantial share of firms hire AI talents from outside their home states, making state-level NCA enforcement regimes plausibly exogenous to firm-level selection or strategic behaviors. Figure 2 shows the geographical distribution of the 3,502 recipient companies in this study. About 400 companies are headquartered in California. Figure 3 shows the state-level employee mobility score in 2023, which is calculated as the opposite of NCA enforcement level following Shi (2023). Given that California's distinctive policy and regulatory environment could systematically influence firms' AI adoption and mobility patterns, we further verify that our instrumental variable estimates are not driven by such regional factors. Appendix B reports the results after excluding all California-based firms, and the overall patterns remain substantially unchanged.

<Figure 2 Inserted Here> <Figure 3 Inserted Here>

The second set of instruments is the Hausman-type instrumental variables, which have been widely adopted in recent IS literature (Wang and Wu, 2024; Zhang et al., 2024, Wu et al., 2018). The basic principle is to use the independent variable of "peer firms" as the instrumental variable for the focal firm's independent variable, under the premise that the hiring behaviors of peer firms are likely to be correlated, but other firms' hiring behavior should not directly impact the output of the focal firm. Specifically, we treat the source companies from which each focal firm hires AI talents in a given year as its "peer firms" and calculate the average AI pool of these source firms in that year as the instrument. The hiring decisions of the focal company and its source companies are likely to be correlated due to shared personnel, while variations in talent seeking by the source companies should not directly impact the focal firm's productivity.

By leveraging two distinct sources of exogenous variation—one at the geographic level and the other at the peer-firm level—we aim to mitigate concerns about endogeneity from any single source. The consistency of results across both instrumental variable strategies would lend greater credibility to the robustness of our estimates. The first-stage F-statistics for all 2SLS results are over the conventional threshold of 20, suggesting that the results do not suffer from weak instrument issues.

To further verify the validity of the instrumental variables used in this study, we follow the procedures in Martin and Yurukoglu (2017) and Wang et al. (2024) to conduct the reduced-form instrumental variable validity test. The logic is as follows: if the instrumental variable is valid, then it should be correlated *only* with the focal company's output performance through its effect on the focal company's independent variable, i.e., AI pool. For companies without an AI pool (i.e., companies that have never hired AI talents from other companies), we should not observe a significant correlation between the instrument and the outcome variable. To test this, we regress the dependent variable on the instrumental variables directly using only data from observations that did or did not have an AI pool. The results are shown in Appendix C. We find that the statistical relationship between the instruments and the dependent variable is significant in samples with an AI pool, whereas it is not in samples without one. These results provide further evidence for the validity of our instrumental variables.

## 5. Results

### 5.1 Main Results

Table 3 reports the OLS regression results. The Cobb-Douglas production function reports the factor share of each component's contribution to the total productivity (Bresnahan, 2003). Column (1) introduces the variable *log AI pool* to illustrate the technological spillover effects stemming from the inflow of AI talent. The coefficient is statistically significant at the 0.01 level. Economically, although the estimated spillover effect of AI on productivity contribution is only around 0.5%, which appears smaller than the 2%-3% productivity contribution typically attributed to IT spillover in the literature (Tambe and Hitt, 2014; Wu et al., 2018), it is important to consider the relative scale of the labor force involved. Prior studies show that IT labor accounts for approximately 2% of the total workforce (Tambe and Hitt, 2012), whereas our data suggest that AI labor currently makes up only about 0.2%. This indicates that, relative to its scale, the spillover effect of AI is substantially larger than that of traditional IT. This result supports Hypothesis H1.

<Table 3 Inserted Here>

Building on this, Columns (2) and (3) separately incorporate *log LSM AI pool* and *log Flat AI pool* to examine how the organizational practices and governance structures of source firms shape the magnitude of these spillovers. Both variables exhibit positive and statistically significant coefficients (0.004 and 0.007, respectively), suggesting that higher levels of LSM practices and flatter organizational structures in the source firms are significantly associated with increased productivity gains through the spillover effect. Note that *log AI pool* is no longer statistically significant, suggesting that AI spillovers are mainly from companies that also adopt LSM or flat structures, whereas hiring from firms lacking these traits yields weaker or no spillover benefits. These findings also validate Hypotheses H2 and H3. Column (4) includes both *log LSM AI pool* and *log Flat AI pool*. In this specification, *log Flat AI pool* remains positively significant ($p < 0.01$), while *log LSM AI pool* becomes statistically insignificant, suggesting that the flat hierarchical structure, associated with higher knowledge diffusion and decentralized decision-making, plays a more dominant role in shaping AI spillovers.

Table 4 shows the instrumental variable 2SLS results. Columns (1) – (3) show the 2SLS results using the state-level enforcement intensity of NCA, and Columns (4) – (6) use Hausman-type instrumental variables. All the results are directionally consistent, supporting the main results. Due to the increases in the coefficients in the 2SLS regressions, we restrict our discussion to the sign of these coefficients, although these could suggest that the magnitude of the effects we observe is in reality larger than what is suggested by the OLS results.

<Table 4 Inserted Here>

One potential concern is that, as an emerging technology, the spillover effect of AI may have been overestimated in recent years due to market hype. To address this, Figure 4 presents the estimated coefficients of AI spillovers and their 95% confidence intervals, based on regressions run separately for each year. As shown in the figure, the magnitude of the AI spillover effect remains relatively stable from 2010 to 2023, with notable increases during 2010–2012 and again after 2021. The former may reflect the wave of foundational investment in AI infrastructure and early talent accumulation, possibly triggered by the breakthroughs in deep learning following the 2009 ImageNet launch. The latter is likely driven by the rapid advancement of generative AI technologies, which have expanded the scope of AI applications across industries.

<Figure 4 Inserted Here>

**5.2 Mechanism Tests**

To verify the proposed mechanism through which the LSM and flat organizational structures enhance the knowledge of AI talent and, in turn, generate higher spillovers, we design four sets of mechanism variables, including one's duty richness, salary changes, product development, and innovation development. If this mechanism holds, employees from high-LSM and flatter organizations should exhibit stronger cross-functional adaptability and demonstrate greater value-creation capacity in both their work and outputs.

<Table 5 Inserted Here>

Under the Lean Startup Model (LSM), firms promote product innovation through rapid prototyping and iterative experimentation, requiring employees to engage across engineering, market validation, and data analysis tasks (Wang and Wu, 2024). Similarly, flatter organizations reduce hierarchical constraints, enabling broader cross-functional participation. Together, high levels of LSM adoption and organizational flatness in source firms foster the development of versatile AI generalists—individuals who accumulate transferable contextual knowledge and generate stronger interfirm spillovers. Columns (1)–(3) of Table 5 use the average duty richness of AI employees as the dependent variable, where a single employee's duty richness is defined as the number of parallel job titles they hold.[9] The results show that both *log LSM AI pool* and *log Flat AI pool* are significantly and positively associated with greater duty richness, indicating that AI talents recruited from LSM-intensive and flat organizations are more likely to be generalists with strong cross-functional collaboration capabilities—skills essential for effective AI deployment (Tambe, 2025).

Furthermore, if AI professionals from flatter or more LSM-intensive source firms generate greater productivity spillovers—as our theoretical framework suggests—their higher marginal value should be reflected in labor market outcomes. Specifically, we expect these individuals to experience larger salary increases upon joining new firms. To test this implication, we construct a salary change indicator based on the difference between their starting salaries at the new firm and their prior compensation levels. Columns (4)–(6) of Table 5 use the average salary change of inflow employees as the dependent variable. The results show that AI talents recruited from LSM-intensive and flat organizations tend to enjoy a higher salary increase in their new positions, consistent with the mechanism.

---

[9] We measure an employee's duty richness by counting the total number of distinct job titles or roles listed in the job title field of Revelio Labs, separated by the symbols ",", "&", "–", "/", or "|". We then compute the average duty richness of all employees hired by the focal firm from its source firms and apply a logarithmic transformation to this value.

Last but not least, if our proposed mechanism holds, then we should observe that firms hiring AI professionals from flatter or more LSM-intensive organizations experience improvements in innovation efficiency. Prior research has highlighted that one of AI's core contributions lies in accelerating product innovation (Babina et al., 2024). Building on this, we examine whether AI talent from high-LSM and flat organizational backgrounds enables recipient firms to iterate on product and technological innovations more rapidly.

To test this mechanism, we measure innovation efficiency using the average time required to obtain trademarks and patents, two complementary indicators of product and technological innovation. Specifically, following Wang and Wu (2024), we construct iteration cycle measures for trademarks and patents, calculated as 365 divided by the total number of trademark applications or granted patents in a given year. Columns (7)–(9) of Table 5 report that both log LSM AI pool and log Flat AI pool are significantly negatively associated with trademark iteration time, suggesting faster product update cycles. Similarly, Columns (10)–(12) show a significant negative relationship between these variables and patent authorization time, indicating that AI talent from flatter and more experimentation-intensive organizations contributes to quicker transformation of R&D outcomes into granted patents.

In summary, our results demonstrate that organizational design—specifically, the adoption of LSM practices and flatter hierarchies—plays a critical role in shaping the capabilities of AI talent. Employees from high-LSM and flat organizations exhibit superior adaptability, higher compensation gains upon job transitions, and a greater ability to accelerate innovation, as reflected in shorter product and patent development cycles. These attributes form a foundational layer of human capital that enhances the effectiveness of AI deployment in recipient firms, leading to higher productivity improvements. Collectively, these findings highlight a transmission channel through which organizational structures shape workforce competencies, thereby driving downstream innovation performance.

**5.3 Robustness Test Controlling General IT Spillover**

A potential concern in the above results is that AI spillover might be correlated with general IT spillover. To further isolate the unique spillover effects of AI, Table 6 introduces the variable *log Non-AI IT pool*, which controls for potential spillovers originating from non-AI IT employees. This measure is constructed in a similar manner to the AI pool: we first identify all employees in IT-related roles[10], exclude those already classified as AI personnel, and then track their inter-firm mobility. Table 6 reports the results.

<Table 6 Inserted Here>

The results show that the coefficient on *log Non-AI IT pool* is consistently positive and statistically significant across all specifications. The scale of the estimated coefficient also aligns with the estimates for IT pools found in prior studies (e.g., Tambe and Hitt, 2014; Wu et al., 2018). Meanwhile, after introducing this control, the coefficient of *log AI pool* remains positive and statistically significant, despite a slight shrinkage in scale. The *log LSM AI pool* and *log Flat AI pool* variables continue to show positive significance in Columns (2) and (3). Column (4) shows that when both log LSM AI pool and log Flat AI pool are included simultaneously, only log Flat AI pool remains significantly positive, while log LSM AI pool becomes insignificant. This result is consistent with the significance and coefficient direction observed in the main regressions.

---

[10] We adopt the employee classification framework from the 2023 policy report *The Race for U.S. Technical Talent* by the Center for Security and Emerging Technology (CSET). This report, based on LinkedIn job data from Revelio Labs, first algorithmically classifies all positions into 1,000 "role_k1000" job clusters and ultimately identifies 111 technical clusters. Our study directly incorporates the complete list of these 111 technical K1000 job clusters as the criterion for defining IT talents. The specific job titles are provided in Table 1.

These findings suggest that, although interrelated, the spillover effects of AI are distinct from those of general IT spillovers. More importantly, the distinction lies not only in the nature of the technologies per se, but also in the complementary organizational environments required for these spillovers to materialize. Traditional IT spillovers are typically enhanced in standardized and scale-driven environments, where codified knowledge and routine processes facilitate the absorption and reuse of IT capabilities, thereby making them complementary to firm size and formalized workflows (Bresnahan et al., 2002; Tambe and Hitt, 2014).

In contrast, AI technologies—particularly those involving machine learning and generative models—thrive in experimental, iterative, and data-rich contexts. The associated spillovers are more likely to take root in flat organizational structures and among firms that adopt LSM practices, which emphasize rapid prototyping, decentralized decision-making, and continuous learning. These environments not only accelerate the integration of tacit AI knowledge but also create feedback loops that amplify productivity gains from incoming AI talent. As such, the organizational complements to AI spillovers differ fundamentally from those of traditional IT, highlighting the need for firms to align their structures and innovation practices with the technological characteristics of AI.

## 6. Discussion

To summarize, this study investigates the inter-firm spillover effects of AI-related labor mobility and provides new empirical evidence—based on data from U.S. publicly listed companies—on how AI technologies diffuse across firms through employee movement and contribute to productivity gains. Unlike the existing literature on IT spillovers, which primarily emphasizes the transfer of technical knowledge, we find that AI spillovers depend not only on the technical expertise carried by talents, but also heavily on the organizational environments from which these talents originate—particularly whether the source firms feature flatter hierarchies and whether they widely adopt the Lean Startup Model (LSM). AI talents from such firms typically undertake a broad range of responsibilities, excel at rapid experimentation and cross-functional communication, and possess stronger problem-solving capabilities and organizational adaptability. These traits make them more effective in driving innovation within new organizations.

Our findings enrich the current understanding of how AI technologies interact with organizational operational mechanisms. Unlike IT systems, which often rely on standardized processes and routinized environments, AI technologies are more dependent on frequent iteration and localized experimentation. Thus, it is overly simplistic to treat AI as a plug-and-play technical tool; rather, it should be viewed as a complex system whose effectiveness depends on alignment with organizational design and operational practices. Firms seeking to unlock the full potential of AI through external hiring must consider not only *who* they hire, but also *from what organizational context* they are hiring. Based on this insight, future research could extend the analysis to AI diffusion through inter-firm partnerships, open-source collaboration, and corporate venture capital, further exploring whether such organizational complementarities apply across different technological domains and industry settings. This would help explain the heterogeneous impact of AI adoption from a more organizational perspective.

Naturally, this study has certain limitations. Our analysis is confined to publicly listed firms in the United States. While these companies are representative in the global AI adoption landscape, differences in labor mobility, organizational forms, and regulatory environments across countries may limit the generalizability of our findings. Future research should validate these patterns using more diverse and international samples. In addition, our study focuses exclusively on knowledge transfer through labor mobility and does not account for other important AI diffusion channels such as inter-organizational collaborations or shared technology platforms. Further studies could incorporate project-level data to explore non-mobility-based mechanisms of AI knowledge dissemination.

From a theoretical perspective, this study responds to recent observations on the uneven returns from AI adoption by identifying the organizational and managerial differences that underlie these variations. We emphasize that, unlike traditional technologies, the effectiveness of AI depends on its alignment with organizational characteristics such as communication structures, tolerance for trial-and-error, and decision-making flexibility. Our findings also contribute to the literature on knowledge spillovers and employee mobility by incorporating variables related to LSM adoption and organizational flatness, thereby identifying the organizational attributes of source firms that shape the spillover potential of AI talent. This extends our understanding of the relationship between human capital and innovation performance.

Practically, this study offers actionable insights for business leaders, talent managers, and policymakers. For firms, recruiting AI talent should not be based solely on algorithmic proficiency or academic credentials; equal attention should be paid to the organizational background and developmental trajectory of the talent. Priority should be given to individuals from flat, experimentation-driven organizations, as they are more likely to contribute effectively to agile, innovation-focused teams. Moreover, firms aiming to absorb and utilize such external AI talent must also undertake internal reforms—flattening organizational hierarchies, granting employees more autonomy for experimentation, and building data-driven, rapid-learning mechanisms. For policymakers, fostering AI diffusion should go beyond technological deployment and involve broader incentives for organizational transformation and managerial innovation. Supporting firms in cultivating learning-oriented and open organizational environments will be essential to maximizing the societal and economic benefits of AI technologies.

**Figure 1: The AI Talent Diffusion Network in 2023**

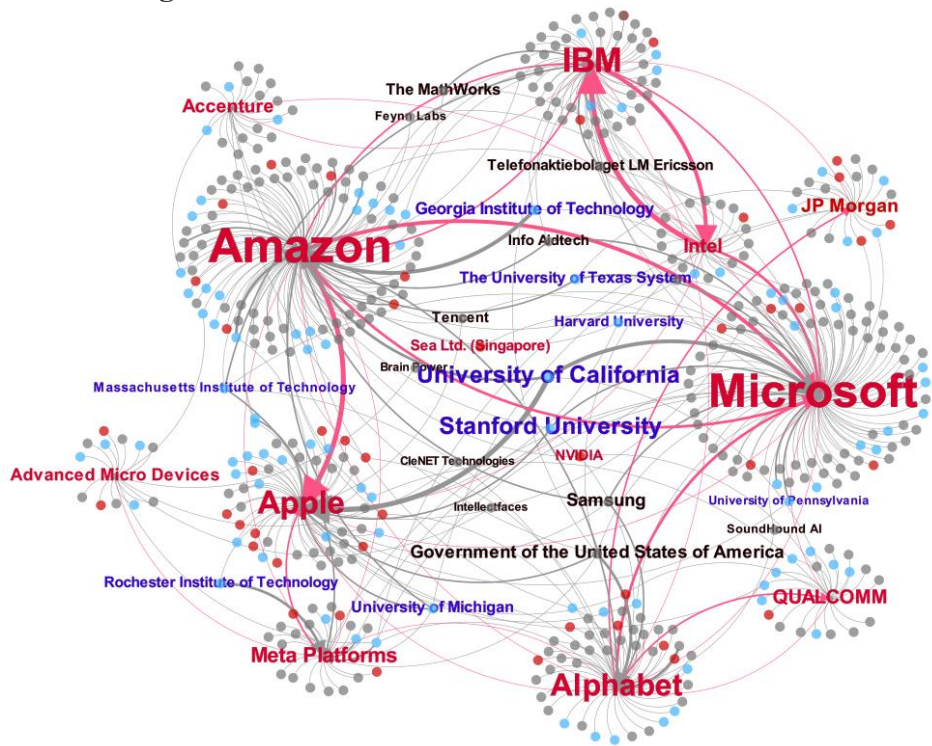

Notes: (1) Node colors represent firm types: red nodes are U.S. publicly listed firms (our sample firms), blue nodes are universities or research institutions, and gray nodes are private companies. Node labels indicate firm names, with font sizes scaled to the number of inbound AI employees. (2) Edge colors denote direction: red edges indicate outbound flows from sample firms, and gray edges indicate inbound flows from other organizations to sample firms. Line width reflects flow frequencies.

**Figure 2: Geographical Distribution of Companies in this Study (All Years)**

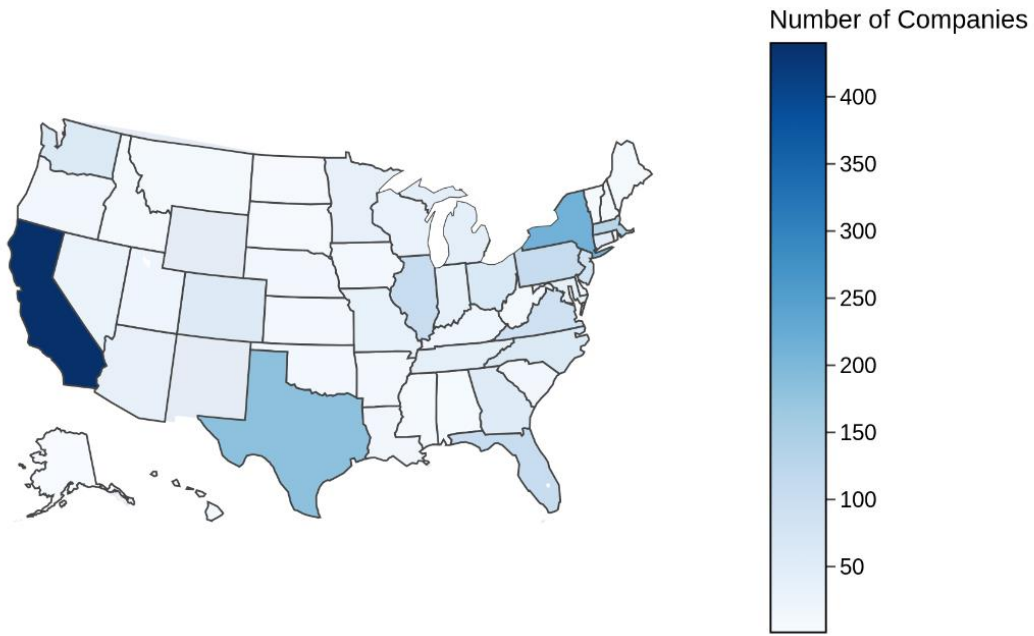

Notes: This plot shows the state-level geographical distribution of the 3,502 recipient public companies in this study. A company's geographical location is determined by its headquarters.

**Figure 3: State-Level Employee Mobility Score (Opposite to NCA Enforcement Level) in 2023**

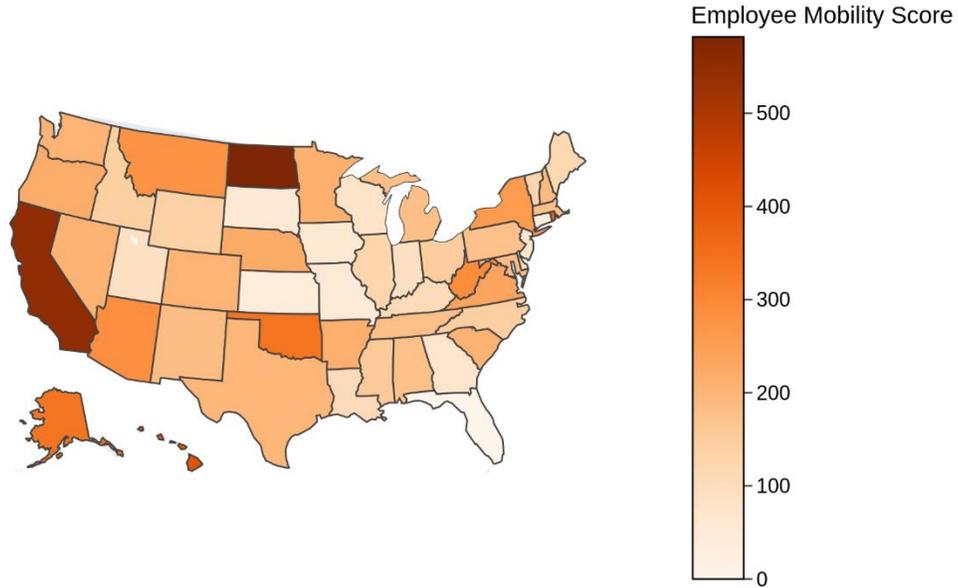

Notes: This plot shows the state-level employee mobility score (opposite to the Non-Compete Agreement enforcement level) in 2023. The measure is calculated following Shi, 2023.

**Figure 4: Coefficient and 95% Confidence Interval of AI Spillover in Different Years**

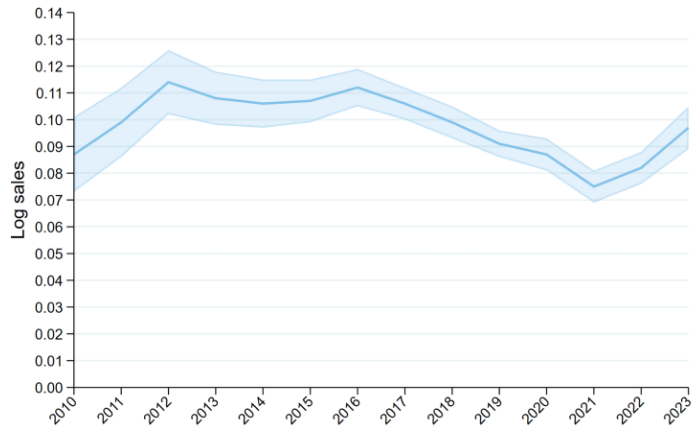

Notes: This plot shows the estimated coefficient (with 95% confidence interval) of AI spillover (*log AI pool*) in each individual year from 2010 to 2023.

# Table 1: Key Words Dictionary

| Topic | Keywords |
|---|---|
| AI | Artificial Intelligence (AI), Machine Learning (ML), Deep Learning (DL), Computer Vision (CV), Neural Network, Convolutional Neural Network (CNN), Recurrent Neural Network (RNN), Natural Language Processing (NLP), Supervised Learning, Unsupervised Learning, XGBoost, TensorFlow, ND4J, Long Short-Term Memory (LSTM), Keras, Autoencoders, Bayesian Networks, Boosting, Caffe Deep Learning Framework, Classification Algorithms, Clustering Algorithms, Computational Linguistics, DBSCAN, Deeplearning4j, Dimensionality Reduction, Dlib, Expectation-Maximization (EM) Algorithm, Gradient Boosting, Hidden Markov Model (HMM), H2O (software), Image Recognition, Information Extraction, JUNG Framework, K-means, Latent Dirichlet Allocation (LDA), Latent Semantic Analysis (LSA), LIBSVM, Matrix Factorization, Maximum Entropy Classifier, Microsoft Cognitive Toolkit (CNTK), MLpack, MXNet, Naive Bayes, Object Recognition, OpenAI Chatbot, AI Chatbot, OpenCV, Opinion Mining, OpenNLP, PyBrain, Random Forests, Recommender Systems, Scikit-learn, SDG (Stochastic Gradient Descent), Sentiment Analysis, Sentiment Classification, Semi-supervised Learning, Support Vector Machines (SVM), Theano, Torch, Vowpal Wabbit, Weka, Word2Vec, Machine Translation, Unstructured Information Management Architecture (UIMA), Apache UIMA, Kernel Methods |
| LSM | MVP (Minimum Viable Product), Test Product, Experimentation, Lab (Technician), (A/B) Testing, Validation, Verification, Trial and Error, Customer/Market Demand, Customer/Market Learning/Response |
| Level of Hierarchies | 1 Owner: (owner, founder, chairman, creator, created, or made)<br>2 CEO: (ceo or any combination of {chief or cheif} and {executive, exec, exective, or exectuiver})<br>3 President: (president or presidente, but not vice)<br>4 CXO: (cco, cdo, cfo, cho, cio, clo, cmo, coo, cpo, cso, cto, or both chief and officer)<br>5 VP: (vp, evp, avp, svp, snrvp, vice president, or vice presidente)<br>6 Head: (head)<br>7 Director: (director, directo, diercto, dir, or dierctor)<br>8 Manager: (manager, mgr, or gm)<br>9 Lead: (lead or leader)<br>10. Supervisor: (supervisor)<br>11. Producer<br>12 Other: (includes none of the above) |
| General IT | .NET Developer, Advisory Software Engineer, Analyst Programmer, Android Developer, Application Development Analyst, Application Development Associate, Cloud Architect, Computer Engineer, Cyber Security Specialist, DBA, Data Architect, Data Center Operator, Data Engineer, Database Administrator, Database Analyst, Database Developer, Database Specialist, Developer, Development Engineer, Development Manager, DevOps Engineer, Digital Product Manager, Embedded Software Engineer, ETL Developer, Frontend Developer, Full Stack Developer, Hardware Engineer, IT Analyst, IT Architect, IT Engineer, IT Operations, IT Project Manager, IT Specialist, Information Analyst, Information Security, Information Security Specialist, Information Systems Specialist, Information Technology Project Manager, Infrastructure Analyst, Infrastructure Architect, Infrastructure Engineer, Integration Engineer, Java Developer, Linux System Administrator, Machine Learning Engineer, Network Analyst, Network Architect, Network Consulting Engineer, Network Engineer, Network Engineering, Network Operations, Network Specialist, Network Support Engineer, Network Technician, Operations Analyst, Oracle Developer, Professor, Programmer, Programmer Analyst, QA Tester, R&D Engineer, R&D Intern, R&D Specialist, SDE, Salesforce Developer, Scientist, Security Analyst, Security Architect, Security Engineer, Share point Developer, Software Consultant, Software Designer, Software Developer, Software Development Engineer in Test, Software Engineer, Software Engineer , Software Engineering, Software Project Manager, |

| | Software Test Engineer, Software Tester, Solutions Architect, Solutions Engineer, Storage Engineer, System Administrator, System Architect, System Developer, System Engineer, Systems Administrator, Systems Analyst, Systems Architect, Systems Engineer, Systems Engineering, Systems Programmer, Technical Architect, Technical Lead, Technical Product Manager, Technical Project Manager, Technical Test Specialist, Technology Lead, Test Automation Engineer, UI Developer, Unix System Administrator, User Experience Researcher, UX Designer |

**Table 2: Descriptive statistics**

| Variable | Obs. | Mean | Std. Dev. | Min | Max |
|---|---|---|---|---|---|
| Sales (mm) | 49028 | 7831.696 | 23489.72 | 0 | 519221.9 |
| Capital (mm) | 49028 | 4064.909 | 14686.91 | 0 | 324117.7 |
| Materials (mm) | 49028 | 4936.119 | 19161.03 | 0 | 484262.7 |
| Advertising Expense (mm) | 49028 | 207.713 | 488.093 | -0.113 | 2299.835 |
| R&D Expense (mm) | 49028 | 288.643 | 1175.986 | -0.648 | 85622 |
| AI labor | 49028 | 16 | 183.577 | 0 | 5944 |
| Non AI labor (m) | 49028 | 21.199 | 63073.69 | 0.001 | 2299.835 |
| AI pool | 49028 | 0.0033 | 0.0226 | 0 | 1 |
| LSM AI Pool | 49028 | 0.0007 | 0.0062 | 0 | 1 |
| Flat AI Pool | 49028 | 1.0527 | 2.8353 | 0 | 12 |

## Table 3: The Effect of AI Spillover on Firm Productivity

| Dependent Variable | Log Sales | | | |
|---|---|---|---|---|
| | (1) | (2) | (3) | (4) |
| Log Capital | 0.326*** | 0.326*** | 0.326*** | 0.326*** |
| | (0.003) | (0.003) | (0.003) | (0.003) |
| Log Materials | 0.498*** | 0.497*** | 0.497*** | 0.497*** |
| | (0.003) | (0.003) | (0.003) | (0.003) |
| Log AI Labor | 0.014** | 0.012** | 0.012* | 0.012* |
| | (0.006) | (0.006) | (0.006) | (0.006) |
| Log non-AI Labor | 0.223*** | 0.223*** | 0.223*** | 0.223*** |
| | (0.003) | (0.003) | (0.003) | (0.003) |
| Log AI Pool | 0.005*** | 0.002 | -0.002 | -0.002 |
| | (0.001) | (0.002) | (0.002) | (0.002) |
| Log LSM AI Pool | | 0.004** | | 0.000 |
| | | (0.002) | | (0.002) |
| Log Flat AI Pool | | | 0.007*** | 0.007*** |
| | | | (0.002) | (0.002) |
| Controls | Year FE, Industry FE, Advertising Expenses, R&D Expenses | | | |
| No. of obs. | 49,027 | 49,027 | 49,027 | 49,027 |
| R-squared | 0.886 | 0.886 | 0.886 | 0.886 |

Notes:

1. A Cobb–Douglas production function model is adopted here.
2. Dependent variable: Log Sales (firm net sales). Inputs: Capital (real fixed assets), Materials (real intermediate inputs excluding labor costs), AI labor (firm-year AI talents identified via job-text keywords), Non-AI labor (total employees at year-end minus AI talents).
3. All variables are log-transformed using $\log(x + c)$, where $c$ equals one half of the smallest positive value of the variable to reduce skewness.
4. Controls: industry fixed effects, year fixed effects, Advertising Expense, R&D Expense.
5. Robust standard errors in parentheses; *** $p<0.01$, ** $p<0.05$, * $p<0.1$.

## Table 4: 2SLS Estimation using Instrumental Variables

| Dependent Variable | Log Sales | | | | | |
|---|---|---|---|---|---|---|
| | (1) | (2) | (3) | (4) | (5) | (6) |
| Log Capital | 0.326*** | 0.326*** | 0.326*** | 0.268*** | 0.267*** | 0.269*** |
| | (0.003) | (0.003) | (0.003) | (0.003) | (0.003) | (0.003) |
| Log Materials | 0.509*** | 0.509*** | 0.509*** | 0.673*** | 0.674*** | 0.672*** |
| | (0.003) | (0.003) | (0.003) | (0.003) | (0.003) | (0.003) |
| Log AI Labor | 0.010 | 0.019* | 0.020* | -0.006 | 0.023*** | 0.007 |
| | (0.012) | (0.011) | (0.011) | (0.017) | (0.008) | (0.012) |
| Log non-AI Labor | 0.218*** | 0.218*** | 0.218*** | 0.169*** | 0.170*** | 0.166*** |
| | (0.004) | (0.004) | (0.004) | (0.004) | (0.004) | (0.004) |
| Log AI Pool | **0.007***** | -0.006 | -0.004 | **0.008**** | -0.015** | -0.076** |
| | **(0.002)** | (0.005) | (0.004) | **(0.004)** | (0.007) | (0.034) |
| Log LSM AI Pool | | 0.011** | | | 0.017** | |
| | | (0.005) | | | (0.007) | |
| Log Flat AI Pool | | | 0.007** | | | 0.077** |
| | | | (0.003) | | | (0.035) |
| Instrumental Variable | Non-Competing Agreements | | | Hausman-type Instrument | | |
| Controls | Year FE, Industry FE, Marketing Expenses, R&D Expenses | | | | | |
| No. of obs. | 49,027 | 49,027 | 49,027 | 49,027 | 49,027 | 49,027 |

Notes:

1. Dependent variable, controls, and fixed effects are the same as Table 3.
2. Columns (1)–(3) use NCA as the instrument variable (state-level non-compete enforceability weighted by the firm's AI inflow source states). Columns (4)–(6) use a Hausman-type instrument variable (the average AI pool of these source firms). The coefficients of instrumented variables are bolded.
3. All specifications report first-stage F > 20, and coefficient signs are consistent with the main results.
4. Robust standard errors in parentheses; *** p<0.01, ** p<0.05, * p<0.1.

## Table 5: Mechanism Tests

| Dependent Variable | Job Duty Richness | | | Salary Change | | | Product Development | | | Innovation Development | | |
|---|---|---|---|---|---|---|---|---|---|---|---|---|
| | (1) | (2) | (3) | (4) | (5) | (6) | (7) | (8) | (9) | (10) | (11) | (12) |
| Log LSM AI Pool | 0.118*** | | -0.003*** | 0.003*** | | 0.003*** | -0.002*** | | 0.000 | -0.029*** | | -0.024*** |
| | (0.000) | | (0.001) | (0.000) | | (0.000) | (0.001) | | (0.002) | (0.002) | | (0.005) |
| Log Flat AI Pool | | 0.132*** | 0.135*** | | 0.002*** | 0.001* | | -0.002*** | -0.002 | | -0.026*** | -0.005 |
| | | (0.000) | (0.001) | | (0.000) | (0.000) | | (0.001) | (0.002) | | (0.002) | (0.005) |
| Controls | Log Capital, Log Materials, Log AI Labor, Log non-AI Labor, Log AI Pool, Year FE, Industry FE, Marketing Expenses, R&D Expenses | | | | | | | | | | | |
| No. of obs. | 49,027 | 49,027 | 49,027 | 49,027 | 49,027 | 49,027 | 49,027 | 49,027 | 49,027 | 49,027 | 49,027 | 49,027 |
| R-squared | 0.859 | 0.940 | 0.940 | 0.027 | 0.025 | 0.027 | 0.011 | 0.011 | 0.011 | 0.384 | 0.383 | 0.384 |

Notes:

1. For Columns (1)–(3), the dependent variable is Job Duty Richness (employee duty richness index); for Columns (4)–(6), the dependent variable is Salary Change (base-salary change for employees switching firms); for Columns (7)–(9), the dependent variable is Product Development (trademark iteration time); and for Columns (10)–(12), the dependent variable is Innovation Development (patent authorization time).

2. Robust standard errors in parentheses; *** $p<0.01$, ** $p<0.05$, * $p<0.1$..

## Table 6: Robustness Test after Controlling the Non-AI IT Pool

| Dependent Variable | Log Sales | | | |
|---|---|---|---|---|
| | (1) | (2) | (3) | (4) |
| Log Capital | 0.332*** | 0.332*** | 0.332*** | 0.332*** |
| | (0.003) | (0.003) | (0.003) | (0.003) |
| Log Materials | 0.494*** | 0.493*** | 0.493*** | 0.493*** |
| | (0.003) | (0.003) | (0.003) | (0.003) |
| Log AI Labor | 0.013** | 0.012** | 0.012** | 0.012* |
| | (0.006) | (0.006) | (0.006) | (0.006) |
| Log non-AI Labor | 0.218*** | 0.218*** | 0.217*** | 0.218*** |
| | (0.003) | (0.003) | (0.003) | (0.003) |
| Log Non-AI IT Pool | 0.005*** | 0.005*** | 0.005*** | 0.005*** |
| | (0.000) | (0.000) | (0.000) | (0.000) |
| Log AI Pool | 0.002** | 0.000 | -0.002 | -0.002 |
| | (0.001) | (0.002) | (0.002) | (0.002) |
| Log LSM AI Pool | | 0.003** | | 0.001 |
| | | (0.001) | | (0.002) |
| Log Flat AI Pool | | | 0.004** | 0.004** |
| | | | (0.002) | (0.002) |
| Controls | Year FE, Industry FE, Marketing Expenses, R&D Expenses | | | |
| No. of obs. | 49,027 | 49,027 | 49,027 | 49,027 |
| R-squared | 0.887 | 0.887 | 0.887 | 0.887 |

Notes:

1. Dependent variable, controls, and fixed effects are the same as Table 3. Each column additionally controls for Log Non-AI IT Pool.
2. Robust standard errors in parentheses; *** p<0.01, ** p<0.05, * p<0.1.)

**Appendix A: Hierarchies in Different Departments**

We adopt 2 methods to identify the within-department hierarchy levels:

**Method 1: Based on the Role k50 variable provided by Revelio Labs**

We use the following keyword dictionary to identify the department to which each employee belongs, then calculate the total number of hierarchies within each department. The number of hierarchical levels is calculated using the method in the main text.

| Department | Keywords in Role 50k |
|---|---|
| Manufacturing | manufactur, production, operator, technician, mechanic, foreman, fabrication, logistic |
| Engineering | engineer, developer, architect, technology, QA, test, data engineer, system, infrastructure, devops |
| Sales and Marketing | sales, marketing, brand, customer, client, retail, business development, account manager |
| Human Resources | human resource, hr, recruit, recruiter, training, talent, people |
| Finance and Accounting | finance, account, auditor, controller, payroll, treasurer, financial, investment, bank |
| Research and Development | research, scientist, laboratory, R&D, chemist, biology, clinical |
| Administrative | admin, assistant, secretary, support, coordinator, operations, procurement, legal, compliance, logistics |

The average hierarchical levels in each department:

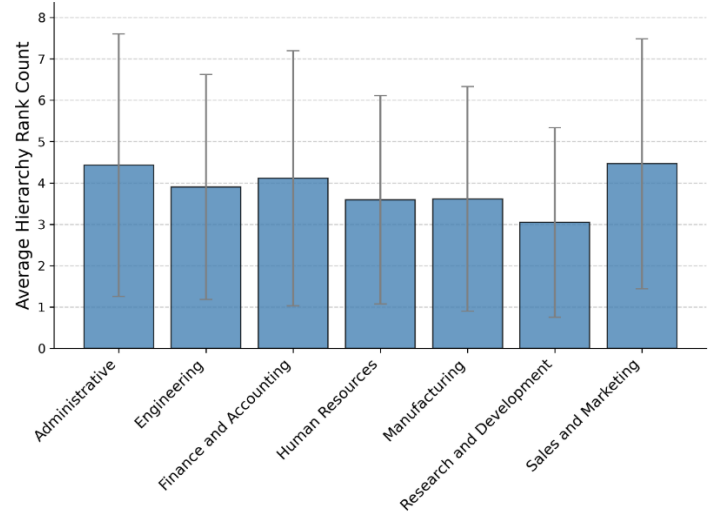

## Method 2: Based on the Job Category variable provided by Revelio Labs

Revelio Labs has also provided the job category of each resume record, which is classified into 7 categories: Admin, Engineer, Finance, Marketing, Operation, Sales, and Scientist. We calculate the total number of hierarchies within each category. The number of hierarchical levels is calculated using the method in the main text. The average hierarchical levels in each department:

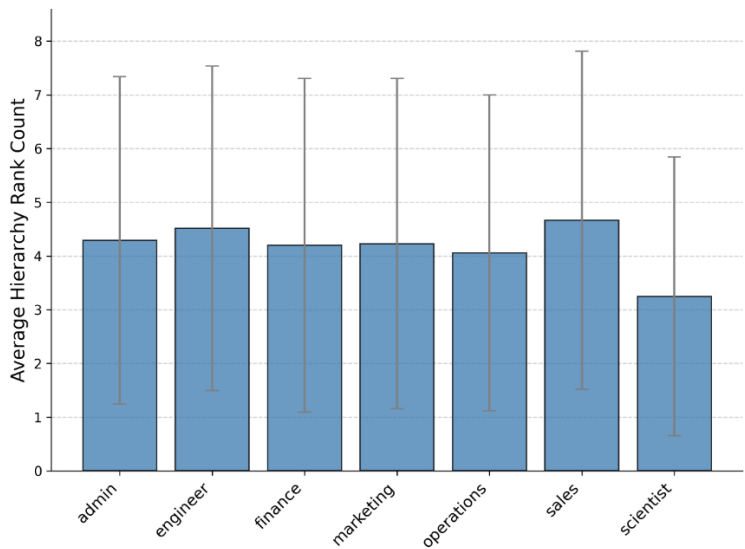

These results indicate that our measure of hierarchical flatness is not substantially biased by departmental heterogeneity within firms.

# Appendix B: Results Excluding California-based Firms

## Table B1: The Effect of AI Spillover on Firm Productivity (Excluding California-based Firms)

| Dependent Variable | Log Sales | | | |
|---|---|---|---|---|
| | (1) | (2) | (3) | (4) |
| Log Capital | 0.303*** | 0.303*** | 0.303*** | 0.303*** |
| | (0.003) | (0.003) | (0.003) | (0.003) |
| Log Materials | 0.528*** | 0.528*** | 0.528*** | 0.528*** |
| | (0.003) | (0.003) | (0.003) | (0.003) |
| Log AI Labor | 0.005 | 0.004 | 0.003 | 0.003 |
| | (0.007) | (0.007) | (0.007) | (0.007) |
| Log non-AI Labor | 0.222*** | 0.222*** | 0.222*** | 0.222*** |
| | (0.004) | (0.004) | (0.004) | (0.004) |
| Log AI Pool | 0.004*** | 0.002 | -0.001 | -0.001 |
| | (0.001) | (0.002) | (0.002) | (0.002) |
| Log LSM AI Pool | | 0.003 | | -0.000 |
| | | (0.002) | | (0.002) |
| Log Flat AI Pool | | | 0.006*** | 0.006*** |
| | | | (0.003) | (0.002) |
| Controls | Year FE, Industry FE, Advertising Expenses, R&D Expenses | | | |
| No. of obs. | 42,867 | 42,867 | 42,867 | 42,867 |
| R-squared | 0.891 | 0.891 | 0.891 | 0.891 |

Notes:

1. A Cobb–Douglas production function model is adopted here.
2. Dependent variable: Log Sales (firm net sales). Inputs: Capital (real fixed assets), Materials (real intermediate inputs excluding labor costs), AI labor (firm-year AI talents identified via job-text keywords), Non-AI labor (total employees at year-end minus AI talents).
3. All variables are log-transformed using $\log(x + c)$, where c equals one half of the smallest positive value of the variable to reduce skewness.
4. Controls: industry fixed effects, year fixed effects, Advertising Expense, R&D Expense.
5. Robust standard errors in parentheses; *** $p<0.01$, ** $p<0.05$, * $p<0.1$.

# Table B2: 2SLS Estimation using Instrumental Variables Productivity (Excluding California-based Firms)

| Dependent Variable | Log Sales | | | | | |
|---|---|---|---|---|---|---|
| | (1) | (2) | (3) | (4) | (5) | (6) |
| Log Capital | 0.305*** | 0.305*** | 0.305*** | 0.303*** | 0.303*** | 0.303*** |
| | (0.003) | (0.003) | (0.003) | (0.003) | (0.003) | (0.003) |
| Log Materials | 0.534*** | 0.534*** | 0.534*** | 0.528*** | 0.528*** | 0.528*** |
| | (0.003) | (0.003) | (0.003) | (0.003) | (0.003) | (0.003) |
| Log AI Labor | 0.008 | 0.017* | 0.018* | 0.004 | 0.004 | 0.001 |
| | (0.013) | (0.011) | (0.018) | (0.007) | (0.012) | (0.012) |
| Log non-AI Labor | 0.218*** | 0.218*** | 0.218*** | 0.222*** | 0.222*** | 0.222*** |
| | (0.004) | (0.004) | (0.004) | (0.004) | (0.004) | (0.004) |
| Log AI Pool | **0.006*** | -0.007 | -0.004 | **0.006*** | 0.002 | -0.007 |
| | **(0.002)** | (0.005) | (0.004) | **(0.004)** | (0.006) | (0.027) |
| Log LSM AI Pool | | **0.011**** | | | **0.009*** | |
| | | **(0.005)** | | | **(0.005)** | |
| Log Flat AI Pool | | | **0.007**** | | | **0.052*** |
| | | | **(0.003)** | | | **(0.028)** |
| Instrumental Variable | Non-Competing Agreements | | | Hausman-type Instrument | | |
| Controls | Year FE, Industry FE, Marketing Expenses, R&D Expenses | | | | | |
| No. of obs. | 42,867 | 42,867 | 42,867 | 42,867 | 42,867 | 42,867 |

Notes:

1. Dependent variable, controls, and fixed effects are the same as Table B1.
2. Columns (1)–(3) use NCA as the instrument variable (state-level non-compete enforceability weighted by the firm's AI inflow source states). Columns (4)–(6) use a Hausman-type instrument variable (the average AI pool of these source firms). The coefficients of instrumented variables are bolded.
3. All specifications report first-stage F > 20, and coefficient signs are consistent with the main results.
4. Robust standard errors in parentheses; *** p<0.01, ** p<0.05, * p<0.1.

**Appendix C: Reduced-Form Instrumental Variable Test**

**Table C1: Reduced-Form Instrumental Variable Validation Test**

| Dependent Variable | Log Sales | | | |
|---|---|---|---|---|
| | (1) | (2) | (3) | (4) |
| | AI_pool≠0 | AI_pool=0 | AI_pool≠0 | AI_pool=0 |
| NCA Enforcement Instrumental Variable | 0.285*** | -0.141 | | |
| | (0.029) | (0.123) | | |
| Hausman-type Instrumental Variable | | | 0.039*** | 0.277 |
| | | | (0.003) | (0.213) |
| Constant | 3.773*** | 2.758*** | 4.527*** | 5.657** |
| | (0.215) | (0.228) | (0.211) | (2.396) |
| Observations | 6,751 | 42,277 | 6,751 | 42,277 |
| R-squared | 0.347 | 0.227 | 0.356 | 0.227 |

Notes:

1. Dependent variable: Log Sales (firm net sales). The independent variables are the 2 sets of instrumental variables used in the main text: (1) the weighted average NCA enforcement level of states from which the focal firm hires AI talent, and (2) the weighted average AI pool level of source companies from which the focal firm hires AI talent.
2. Columns (1) and (3) show regression results using only the sample of observations that have never hired AI talents before. Columns (2) and (4) show regression results using the rest of the observations. The results show that both instrumental variables are significantly correlated with the dependent variables only when the focal firm has acquired AI talent, suggesting that they correlate with the original dependent variable only through the original independent variable.
3. Robust standard errors in parentheses; *** p<0.01, ** p<0.05, * p<0.1.